\documentclass{aa}  

\usepackage{graphicx}
\usepackage{txfonts}
\usepackage{natbib,twoopt}
\usepackage{amsmath}
\usepackage[nopatch]{microtype}
\usepackage{booktabs}
\usepackage{multirow}
\usepackage{hyphenat}
\usepackage{ulem}
\usepackage{amssymb}
\usepackage{threeparttable}
\usepackage{color,soul}

\usepackage{xcolor}
\usepackage[colorlinks=True,allcolors=blue]{hyperref}

\newcommand{\eros}{{\it eROSITA}}
\newcommand{\sw}{{\it Swift}}
\newcommand{\fermi}{{\it Fermi}}
\newcommand{\nh}{$N_{\rm H}$} 

\newcommand{\fgl}{J1824} 
\newcommand{\fgljII}{4FGL~J1824.2+1231} 

\newcommand{\ergs}{erg~s$^{-1}$}
\newcommand{\ergscm}{erg~s$^{-1}$~cm$^{-2}$}
\newcommand{\msun}{M$_\odot$}
\def\degs{\ifmmode ^{\circ}\else$^{\circ}$\fi}
\def\amin{\ifmmode ^{\prime}\else$^{\prime}$\fi}
\def\asec{\ifmmode ^{\prime\prime}\else$^{\prime\prime}$\fi}
\def\fss{\hbox{$.\!\!^{\rm s}$}}        
\def\farcs{\hbox{$.\!\!^{\prime\prime}$}}  
  
\def\h{$^{\rm h}$}
\def\m{$^{\rm m}$}

\newcommand{\flux}{erg~s$^{-1}$~cm$^{-2}$} 
\newcommand{\fluxd}{erg~s$^{-1}$~cm$^{-2}$~\AA$^{-1}$}
\def\gaia{\textit{Gaia}}
\def\ps{Pan-STARRS}
\def\gem{\textit{Gemini}}

\begin{document} 

\titlerunning{Is 4FGL~J1824.2+1231 a tMSP?} 
\authorrunning{Zyuzin et al.}

\title{Is the \fermi\ source 4FGL~J1824.2+1231 \\ a transitional millisecond pulsar?}
  
\author{
D. A. Zyuzin\inst{1}\thanks{da.zyuzin@gmail.com},
A. V. Karpova\inst{1},
A. Yu. Kirichenko\inst{2,1},  
Yu. A. Shibanov\inst{1}, 
I. F. Bikmaev\inst{3,4}, 
M. R. Gilfanov\inst{5,6}, \\
E. N. Irtuganov\inst{3},
M. A. Gorbachev\inst{3},
M. V. Suslikov\inst{3},
R. Karimov\inst{7},
M. M. Veryazov\inst{8}, 
M. Pereyra\inst{9}
}

\institute{
Ioffe Institute, 26 Politekhnicheskaya, St. Petersburg, 194021,  Russia 
\and         
Instituto de Astronom\'ia, Universidad Nacional Aut\'onoma de M\'exico, Apdo. Postal 106, Baja California, M\'exico, 22860               
\and
Department of Astronomy and Satellite Geodesy, Kazan Federal University, Kremlevskaya Str. 18, Kazan, 420008, Russia
\and Academy of Sciences of Tatarstan, Baumana Str. 20, 420111 Kazan, Russia
\and Space Research Institute of the Russian Academy of Sciences, Profsoyuznaya Str. 84/32, 117997 Moscow, Russia 
\and Max-Planck-Institut für Astrophysik, Karl-Schwarzschild-Str. 1, D-85741 Garching, Germany
\and Ulugh Beg Astronomical Institute, Uzbekistan Academy of Sciences, Tashkent, 100052, Uzbekistan
\and Peter the Great St. Petersburg Polytechnic University,  Polytechnicheskaya st.,
29, St. Petersburg, 195251, Russia
\and SECIHTI, Instituto de Astronom{\'\i}a, Universidad Nacional Aut\'onoma de M\'exico, 22860 Ensenada, BC, Mexico}

\date{Received ..., 2025; accepted ...}
 
\abstract
{Transitional millisecond pulsars (tMSPs) in tight binary systems 
represent an important  evolutionary  link 
between low-mass X-ray binaries and radio millisecond pulsars. 
To date, only three confirmed tMSPs and a few candidates have been discovered. 
Most of them are $\gamma$-ray sources.   
For this reason, searching for multiwavelength counterparts to unassociated  \fermi\ $\gamma$-ray sources can help to find new tMSPs.}  
{Here we investigate whether the unassociated $\gamma$-ray source 4FGL~J1824.2+1231 belongs to the tMSP family.}
{To find the counterpart to 4FGL~J1824.2+1231, we used data from SRG/\eros\ and \sw\ X-ray catalogues,  
and from different optical catalogues. We also performed 
time-series photometric optical observations of the source with the 2.1-m telescope of the Observatorio Astron\'omico Nacional San Pedro M\'artir, the 1.5-m telescope of the Maidanak Astronomical Observatory and the 1.5-m Russian-Turkish telescope.
In addition, we carried out optical spectroscopic observations with the Russian-Turkish telescope and used archival spectroscopic data obtained with the \gem-North telescope.  
}
{Within the position error ellipse of 4FGL~J1824.2+1231, we found only one X-ray source which coincides with an optical object. 
We consider it as a likely multiwavelength counterpart to 4FGL J1824.2+1231.
The source shows strong optical variability and significant proper motion.
The latter strongly implies that this is a Galactic source.  
Double-peaked H and He emission lines are detected in its spectrum with a flat continuum, 
as often observed in accretion disks of compact binary systems. 
The X-ray spectrum is well fitted by a power law with the photon index  $\sim$1.7.  
The derived intrinsic X-ray-to-$\gamma$-ray flux ratio is about 0.2.}
{If the X-ray/optical source is the true counterpart to 4FGL J1824.2+1231, then all its properties suggest that it is a tMSP in the subluminous disk state.}


\maketitle
%

\section{Introduction}

The \fermi\ Gamma-ray Space Telescope has identified a large number of binary millisecond pulsars (MSPs) including members of the `spider' class, redbacks (RBs) and black widows.
(BWs; e.g. \citealt{roberts2013,strader2021,swihart2022}).
These are compact binaries with orbital periods $P_b \lesssim 1$~d in which one side of the companion star is heated by the pulsar wind.  
RBs have companions with masses of 0.1--1~\msun, while masses of BW companions are $\lesssim$0.05~\msun. It is commonly accepted that MSPs are formed in binary systems where they are spun-up to short rotational periods ($P<30$ ms) due to accretion of matter and angular momentum from their main-sequence companions during the low- or intermediate-mass X-ray binary stages \citep{Bisnovatyi-Kogan1974,alpar1982}. 
Of particular interest is the discovery of RBs that show transitions between the rotation-powered and active X-ray 
regimes, confirming the tight evolutionary link between low-mass X-ray binaries (LMXBs) and radio MSPs (see \citealt{papitto&demartino2022} and references therein).

As the companion in spider binaries is ablated by the energetic pulsar wind, the ablated material creates a cloud around the pulsar, obscuring the pulsar radio emission and causing eclipses. For this reason, detection of radio pulsations from spiders can be challenging.  
However, such objects can be identified through X-ray and optical observations of unassociated \fermi\ sources, even when pulsations are not detected, through periodic modulations of the optical and/or X-ray flux caused by the orbital motion in a binary system \citep[e.g.][]{salvetti2017}.

To date, only three confirmed transitional MSPs (tMSPs) have been detected, namely, PSRs J1023+0038 \citep{Archibald2009}, J1227$-$4853 \citep{Bassa2014,Roy2015} and J1824$-$2452I \citep{Papitto2013}. In addition, several sources were proposed as candidate tMSPs in the so-called subluminous disk state \citep[e.g.][]{bogdanov2016, strader2016, cotizelati2019, miller2020} considering their spectral properties: H and He emission lines in the optical due to the presence of the accretion disk, power-law-shaped spectrum in X-rays with the photon index $\approx 1.7$, and variability such as flaring activity, high/low brightness modes in the ultraviolet (UV), optical and X-rays. 
The $\gamma$-ray emission in the sub-luminous state is a few times brighter than in the rotation-powered regime \citep[][and references therein]{papitto&demartino2022}. 
In the radio, continuum emission with a flat or a slightly inverted spectrum is observed. 
For PSR J1023+0038, simultaneous radio and X-ray observations revealed an anti-correlated variability pattern: the radio flux increases when the source switches from the high to the low X-ray mode \citep{bogdanov2018}.

Transitional MSPs are of great interest for studying the binaries' evolution, accretion processes, interaction of the pulsar magnetic field and pulsar wind with in-flowing plasma, and fundamental physical properties of superdense matter of neutron stars (NSs). 
Thus, new identifications and studies of companions in tMSPs are necessary. 
In this regard, investigations of about 2600 unassociated \fermi\ sources are promising: as suggested in recent studies, 
$\sim$8\% of these sources are good binary or isolated pulsar candidates \citep{kerby2021,mayer&becker2024}.

Here we report on the discovery of the likely X-ray/optical counterpart to the unassociated $\gamma$-ray source \fgljII\ (hereafter \fgl). 
According to the \fermi\ Large Area Telescope 
14-Year Point Source Catalog \citep{4fgl-dr4}, its flux in the 0.1--100 GeV band is $F_\gamma = 2.7(9) \times 10^{-12}$ \flux,
its $\gamma$-ray spectrum 
can be described by the LogParabola model,
and no significant variability was detected, 
 implying that it could be a pulsar.
We found that  the \fgl\ 
counterpart candidate demonstrates properties 
typical of a tMSP in the subluminous accreation disk state.
The X-ray/optical identification is described in Sec.~\ref{sec:counterpart} and observations -- in Sec.~\ref{sec:data}.
Analysis and results are presented in Sec.~\ref{sec:results},
and discussion and conclusions are given in Sec.~\ref{sec:discussion}.
\begin{figure*}
\begin{center}
\begin{minipage}{0.49\linewidth}
\includegraphics[width=1.0\linewidth, trim={0.0cm 0.6cm 0.7cm 0.0cm}, clip]{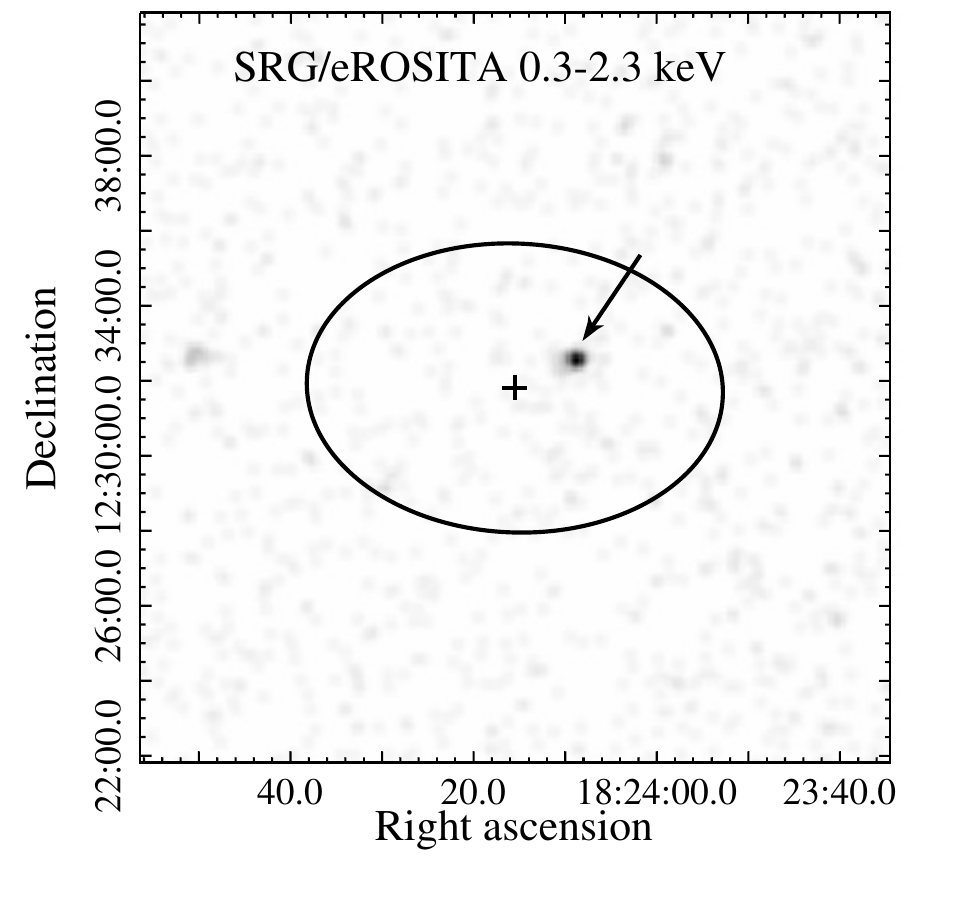}
\end{minipage}
\begin{minipage}{0.49\linewidth}
\includegraphics[width=1.0\linewidth, trim={0.4cm 0.4cm 0.4cm 0.4cm}, clip]{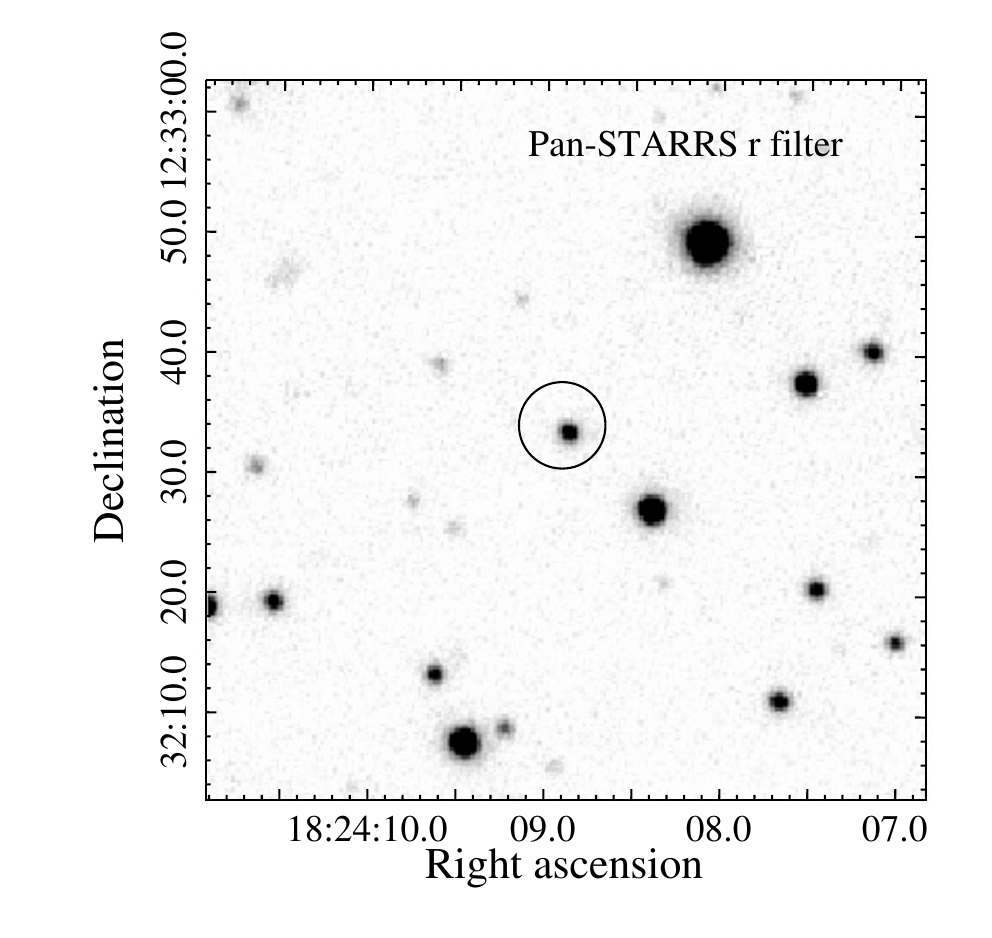}
\end{minipage}
\end{center}
\caption{Images of the \fgl\ field. {\it Left}: 20\amin\ $\times$ 20\amin\ \eros\ image in the 0.3--2.3 keV range.  
The \fgl\ $\gamma$-ray position is marked by the cross while the ellipse shows the 68\% position uncertainty.
The likely X-ray counterpart of \fgl\ is marked by the arrow.
{\it Right}: 1\amin\ $\times$ 1\amin\ \ps\ image in the $r$ band. 
The circle shows the 98\% position uncertainty of the X-ray source obtained with \eros. 
The likely optical counterpart is seen inside the circle. }
\label{fig:xrays}
\end{figure*}
\begin{figure*}
\begin{center}
\begin{minipage}{1.0\linewidth}
\includegraphics[width=1.0\linewidth]{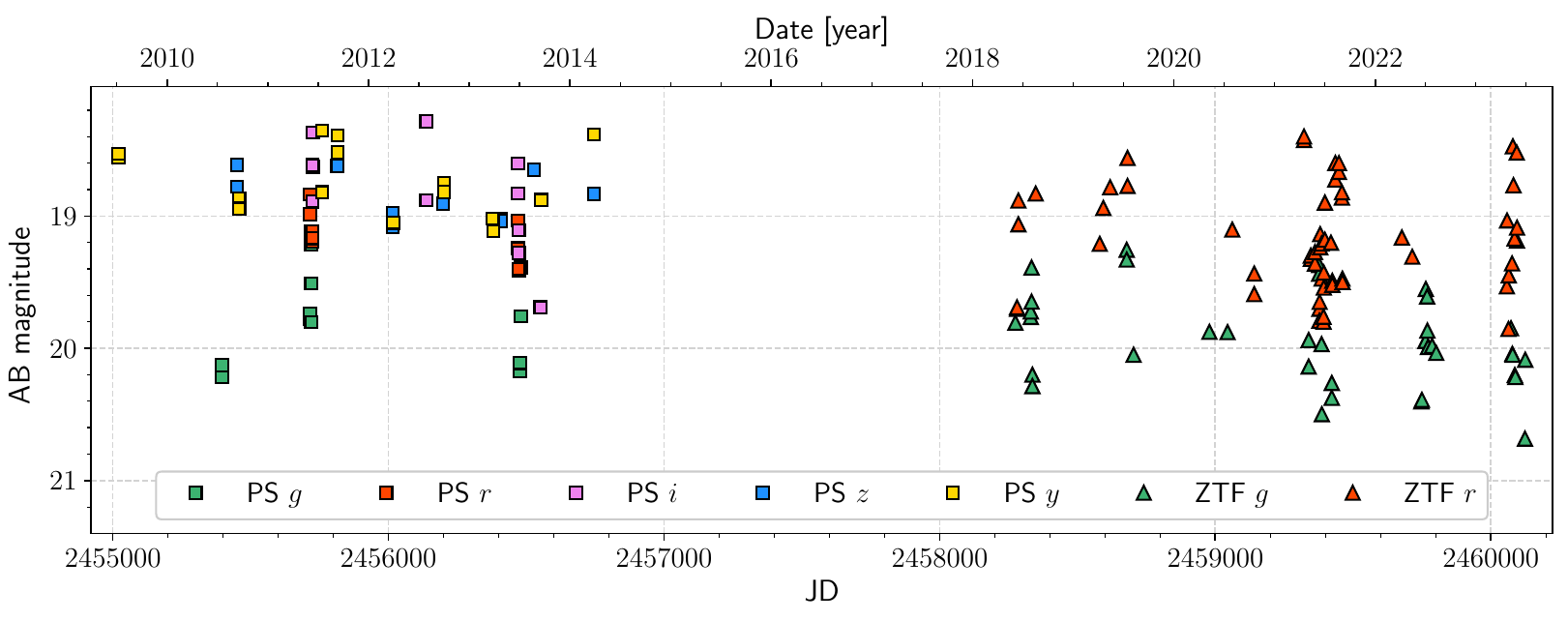}
\end{minipage}
\end{center}
\caption{Optical light curves of the \fgl\ counterpart candidate based on the data from the \ps\ (PS) and ZTF catalogs in different filters indicated in the legend.}
\label{fig:lc-cat}
\end{figure*}

\section{X-ray/optical counterpart identification}
\label{sec:counterpart}

We examined X-ray data of the \fgl\ field obtained with the extended ROentgen Survey with an Imaging Telescope Array (\eros) telescope \citep{erosita2021} aboard the Spectrum-RG (SRG) orbital observatory \citep{Sunyaev2021}. In addition, we revised the Living \sw-XRT Point Source (LSXPS; \citealt{evans2023}) catalogue. 
In both surveys, we found only one X-ray source inside the \fermi\ 68\% confidence error ellipse corresponding to \fgl\ 
(see the left panel of Fig.~\ref{fig:xrays}). 
For \eros, the designation of the latter is SRGe J182408.9+123234, while for LSXPS it is J182408.7+123231. 
Its coordinates are R.A. = 18\h24\m08\fss91 and Dec. = +12\degs32\amin34\farcs1 with the 98\% position uncertainty of 3\farcs6. 
We propose the source as an X-ray counterpart candidate to the $\gamma$-ray object. 

Investigating optical data from the Panoramic Survey Telescope and Rapid Response System survey Data Release~2 \citep[\ps\ DR~2;][]{flewelling2020}, 
we found a source, PSO~J276.0370+12.5426, that spatially coincides with the \fgl\ X-ray counterpart candidate (see Fig.~\ref{fig:xrays}, right) 
and thus can be considered as its optical counterpart.
Its coordinates are R.A. = 18\h24\m08\fss88660(15) 
and Dec = +12\degs32\amin33\farcs400(2). 
The source is also presented in the \gaia\ DR~3 \citep{gaia2016,gaia2023} and Zwicky Transient Facility DR~23 \citep[ZTF;][]{ztf2019} catalogues. 
According to {\it Gaia}, 
the distance to the source is in the range 1.3--3.8 kpc \citep{bailer-jones2021} and its proper motion is $\mu=18.5(3)$ mas~yr$^{-1}$, implying the Galactic origin.

The ZTF and \ps\ data demonstrate a strong variability  
of the optical candidate at time-scales from  
a month 
to a couple of years, 
with the brightness in the $r$-band changing from 18.4 to 19.9 mag (see Fig.~\ref{fig:lc-cat}). 
However, the small number of optical observations does not allow us to perform timing analysis to search for possible periodic variations. 
In addition, the source was detected in the UV with the Galaxy Evolution Explorer (GALEX) \citep{GALEX}.
Its observed flux density in the Near-UV (NUV; $\sim$2310~\AA) band is $(4.2 \pm 1.5)\times10^{-17}$ \fluxd.
The object was also detected in the infrared with WISE in two bands: 
W1 (3.4 $\mu$m) = 276.5$\pm$7.6 and 	
W2 (4.6 $\mu$m) = 428.9$\pm$26.6	
(all fluxes are in Vega nanomaggies \citep{nanomaggies}).
This corresponds to flux densities of  
W1 = $2.30(6)\times10^{-18}$ \fluxd\ and  W2=$1.02(6)\times10^{-18}$ \fluxd.

To shed light on the nature of the optical/X-ray
source, we reanalysed the X-ray  data and performed
new multi-epoch 
optical 
spectroscopic and time-series photometric
observations.


\section{Observations}
\label{sec:data}


\subsection{Optical photometry and spectroscopy}
\label{subsec:photometry}

Time-series photometric 
optical 
observations of  \fgl\ were performed with the 2.1-m telescope at the Observatorio Astron\'omico Nacional San Pedro M\'artir (OAN-SPM) in Mexico, 1.5-m telescope at the Maidanak Astronomical Observatory (MAO) in Uzbekistan and the Russian-Turkish 1.5-m optical telescope, RTT-150, Turkish National Observatories, Observatory in Antalya, Turkey\footnote{\url{https://tug.tubitak.gov.tr/en/teleskoplar/rtt150-telescope-0}}.
RTT-150 was also used for optical spectroscopy. 
The log of observations is given in Tables~\ref{log1}  and ~\ref{log2}. 


\begin{table}
\caption{Log of  photometric observations of the \fgl\ likely counterpart with the OAN-SPM 2.1-m, Maidanak 1.5-m and RTT-150 telescopes.} 
\label{log1}
\begin{tabular}{cccc}\hline
Date       & JD    & Band & Frames $\times$ exposure time, \\ 
yy/mm/dd   &       &      & s   \\
\hline
\multicolumn{4}{c}{\textbf{OAN-SPM 2.1-m} } \\ 
 2023/05/22 & 2460086 & $R$    &  30 $\times$ 150      \\
 2023/05/23 & 2460087 & $B$    &  21 $\times$ 150     \\  
 2023/05/23 & 2460087 & $R$    &  71 $\times$ 150     \\ 
 2023/05/24 & 2460088 & $B$    &  12 $\times$ 150     \\ 
 2023/05/24 & 2460088 & $R$    &  12 $\times$ 150     \\ 
 2023/05/25 & 2460089 & $B$    &  35 $\times$ 150     \\ 
 2023/05/25 & 2460089 & $R$    &  35 $\times$ 150     \\  
 2023/05/31 & 2460095 & $V$    &  44 $\times$ 150      \\
 2023/06/01 & 2460096 & $V$    &  48 $\times$ 150       \\
\hline
 2023/09/13 & 2460201 & $V$    &  11 $\times$ 300       \\
 2023/09/14 & 2460202 & $V$    &  32 $\times$ 300       \\
 2023/09/15 & 2460203 & $V$    &  36 $\times$ 300       \\
 2023/09/16 & 2460204 & $V$    &  31 $\times$ 300      \\
 2023/09/17 & 2460205 & $V$    &  25 $\times$ 300       \\
\hline
\multicolumn{4}{c}{\textbf{MAO 1.5-m}} \\ 
 2023/09/13 & 2460201 & $V$    &  27 $\times$ 300       \\
\hline
\multicolumn{4}{c}{\textbf{RTT-150}} \\ 
2022/09/13 & 2459836 & $r$     & 37 $\times$ 300  \\
2024/07/26 & 2460518 & $G$     & 48 $\times$ 180   \\
\hline
\end{tabular}
\end{table}

\begin{table}
\caption{Log of  optical spectroscopic observations of the \fgl\ likely counterpart with the  RTT-150 telescope.} 
\label{log2}
\begin{tabular}{cccc}\hline
Date       & JD     & Frames $\times$ exposure time,   \\ 
yy/mm/dd   &        & s   \\
\hline
2022/09/10 & 2459833            & 2 $\times$ 3600      \\
2024/07/23 & 2460515             & 3 $\times$ 3600     \\
2024/07/24 & 2460516             & 2 $\times$ 3600   \\
2024/07/27 & 2460519              & 2 $\times$ 3600     \\
2024/08/10 & 2460533             & 2 $\times$ 3600    \\
2024/08/12 & 2460535             & 2 $\times$ 3600     \\
\hline
\end{tabular}
\end{table}


\subsubsection{OAN-SPM photometry}

\label{OAN-SPM}

The OAN-SPM observations were carried out using the $B$, 
$V$ and $R$ 
Johnson-Cousins filters with the `Rueda Italiana' instrument\footnote{\url{https://www.astrossp.unam.mx/en/users/instruments/ccd-imaging/filter-wheel-italiana}} 
in May, June and September 2023. 
The field of view (FoV) of the detector is 6\amin~$\times$~6\amin\ with an image scale of 0\farcs34 in the 2$\times$2 CCD pixel binning mode. 
The photometric calibration was performed using the Landolt standards SA 109-949, 954, 956  \citep{1992AJ....104..340L}.


\subsubsection{MAO photometry}
The MAO observations of J1824 were carried out on September 13, 2023. 
We used the AZT-22 1.5-m telescope (f/7.74) with the ANDOR- iKon-XL 4k CCD and Bessell \textit{V} filter. The detector pixel scale is 0\farcs266 pix$^{-1}$, 
and the field of view is about 18\farcm1~$\times$~18\farcm1.  
A total of 27 frames were taken during about 3.5 hours with an exposure time of 300 s.


\begin{figure*}[h]
\begin{center}
\begin{minipage}{0.49\linewidth}
\includegraphics[width=1.0\linewidth, clip]{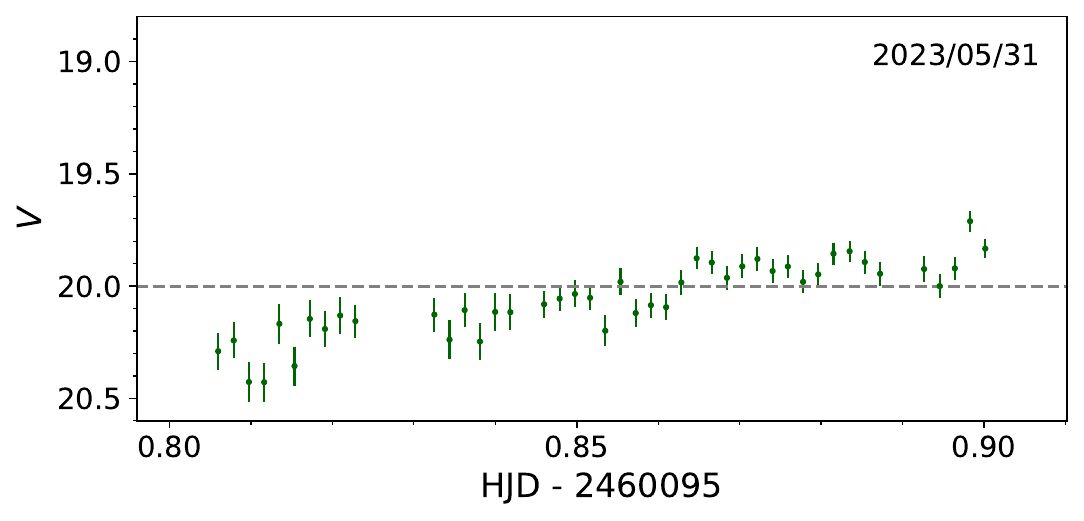}
\end{minipage}
\begin{minipage}{0.49\linewidth}
\includegraphics[width=1.0\linewidth, clip]{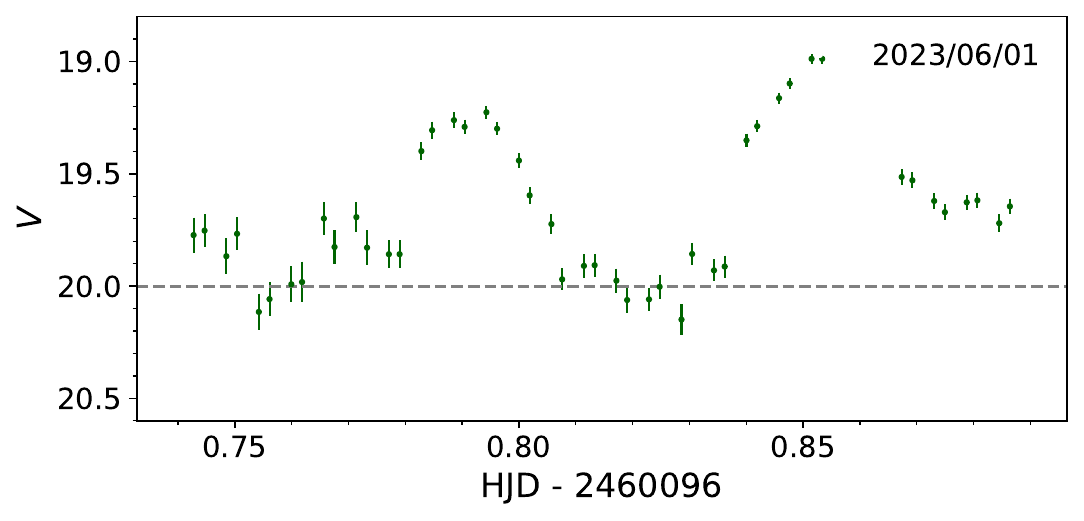}
\end{minipage}
\begin{minipage}{0.49\linewidth}
\includegraphics[width=1.0\linewidth, clip]{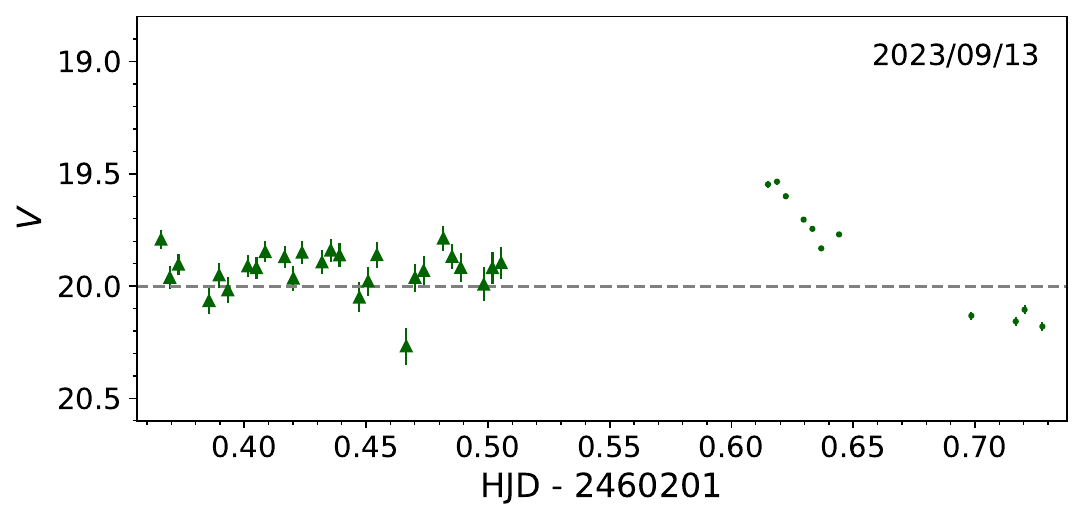}
\end{minipage}
\begin{minipage}{0.49\linewidth}
\includegraphics[width=1.0\linewidth, clip]{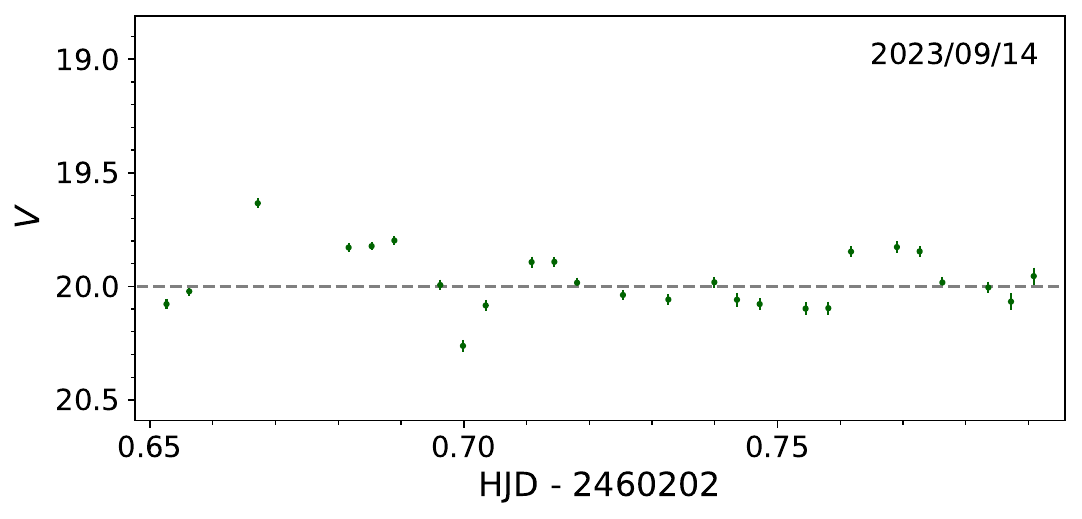}
\end{minipage}
\begin{minipage}{0.49\linewidth}
\includegraphics[width=1.0\linewidth, clip]{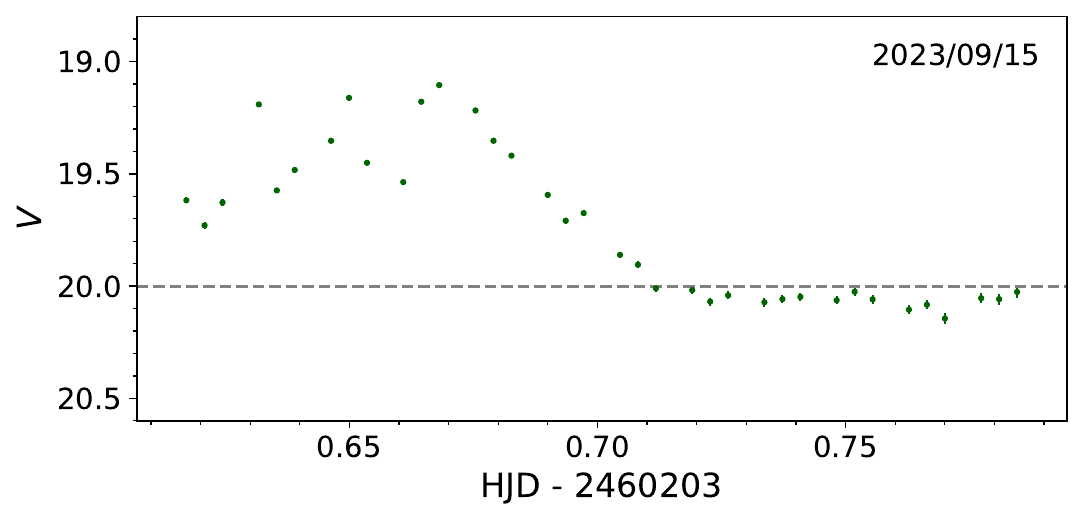}
\end{minipage}
\begin{minipage}{0.49\linewidth}
\includegraphics[width=1.0\linewidth, clip]{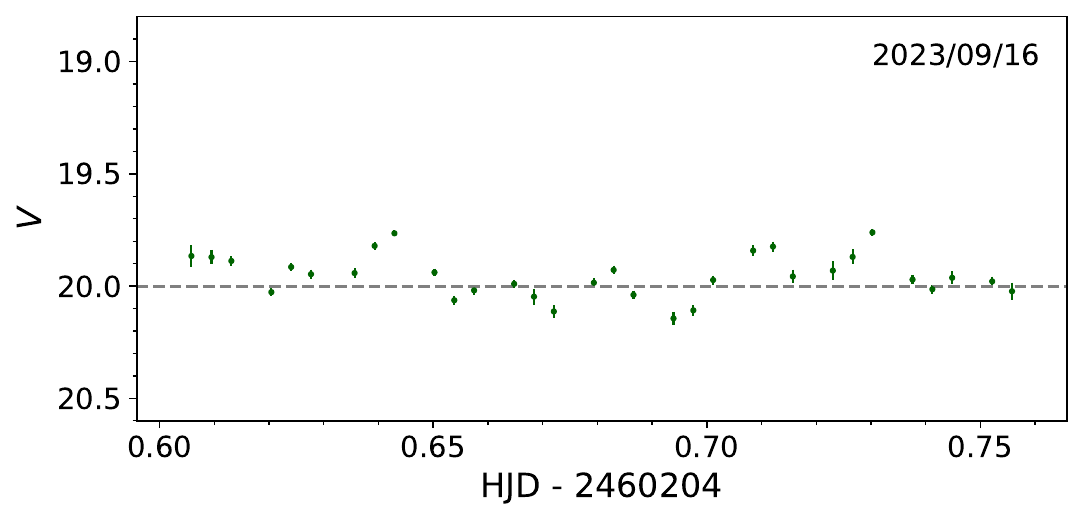}
\end{minipage}
\end{center}
\caption{Heliocentric light curves of the likely optical companion to \fgl\ obtained with the OAN-SPM (dots) and Maidanak (triangles) telescopes in the $V$ band on different dates indicated in the panels.
The dashed line shows the level of the possible stable state of the source ($V = 20.0$ mag) which is the median magnitude defined excluding flaring episodes.}
\label{fig:opt-lc-V}
\end{figure*}
\subsubsection{RTT-150 photometry}
 
The RTT-150 optical photometry was performed using the $r$ filter on September 13, 2022.  
To increase the time resolution to 3 min, observations without filters were performed on July 26, 2024, with the passband close to the \gaia\ $G$-band. 
The telescope is equipped with the TFOSC instrument with the Andor iKon-L 936 BEX2-DD-9ZQ 2048 $\times$ 2048-pixel CCD array.
The TFOSC field of view is 11\amin$\times$11\amin\ with a pixel scale of 0\farcs33 for the 1$\times$1 binning.


\subsubsection{RTT-150 optical spectroscopy}
\label{subsec:opt-spec}

We obtained low-resolution optical spectra of the source with 
the RTT-150 + TFOSC between 2022 and 2024 
(see Table~\ref{log2}). 
Couple spectra with an exposure time of 3600 s were obtained using grism 15 and a 134$\mu$ entrance slit (corresponding to 2\farcs4 at the sky).
The wavelength range was 3900--8900 \r{A} and the spectral resolution was 15 \r{A}.

\subsubsection{Archival optical spectroscopy with \gem-N}

Spectral observations of the optical counterpart were carried out with the \gem-North telescope on August 19, 2022 (PI S. Swihart). 
Two consecutive long-slit spectra with 15-min exposures were obtained using the GMOS instrument with the R400\_G5305 grating in conjunction with the 
GG455\_G0305 longpass filter  
and the slit width of 1\asec, resulting in the spectral coverage 
4500--10300 \AA\  and  resolution
of 8 \AA\ at a blaze wavelength of 7640 \AA.


\subsection{X-ray data}
\label{subsec:x-ray-data}

\fgl\ field was observed in X-rays with \sw\ X-Ray Telescope (XRT) eight times in 2019--2020 with the total exposure time of 3.6 ks (ObsIDs 03110651001, 03110651002, 03110651005, 03110651007--03110651009). 

In 2019--2021, the \fgl\ field was  observed  with 
\eros\ with a total exposure was 840 s (vignetting corrected exposure -- 509 s).


\section{Analysis and results}
\label{sec:results}


\subsection{Optical light curves at short time-scales}
\label{subsec:opt-lc}  

Using the Image Reduction Analysis Facility (IRAF) package \citep{tody1986,tody1993}, we carried out standard processing of the photometric data, including bias subtraction and flat-fielding. We employed the aperture photometry to extract the target light curves. Using a differential technique with the non-variable bright field star \gaia\ DR~3 4485064648767922944\footnote{Its ICRS coordinates are R.A.(J2016) = 18\h24\m08\fss419 and Dec.(J2016) = +12\degs32\amin26\farcs97.} as a reference, we eliminated the variations caused by unstable weather conditions during the observations. 

\begin{figure}[h]
\begin{center}
\begin{minipage}{1\linewidth}
\includegraphics[width=1.0\linewidth, clip]{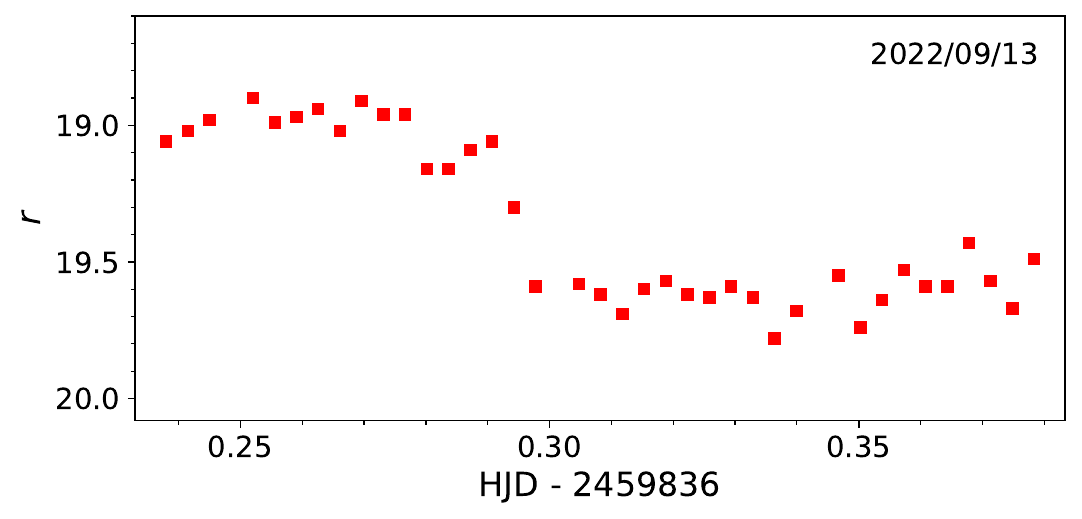}
\end{minipage}
\begin{minipage}{1\linewidth}
\includegraphics[width=1.0\linewidth, clip]{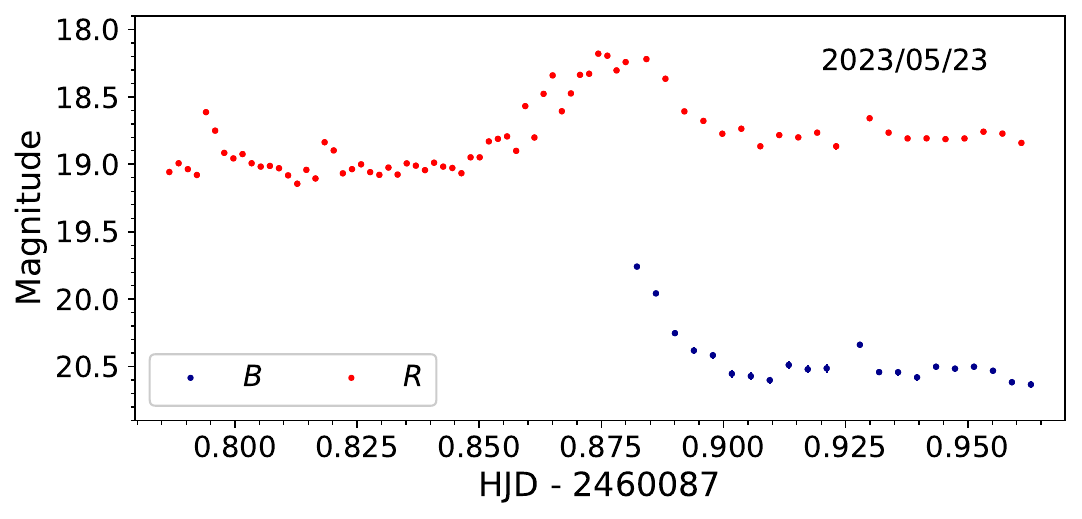}
\end{minipage}
\begin{minipage}{1\linewidth}
\includegraphics[width=1.0\linewidth, clip]{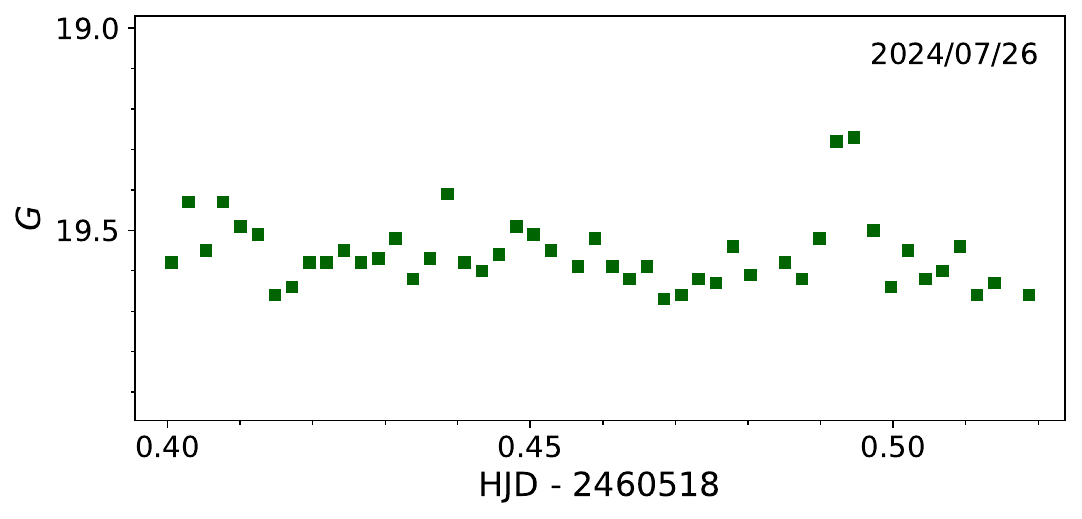}
\end{minipage}
\end{center}
\caption{Heliocentric light curves of the likely optical companion to \fgl\ obtained with the RTT-150 (squares; top and bottom panels) and OAN-SPM (dots; middle panel) telescopes on different dates indicated in the panels. $G$, $B$ and $R$-band magnitudes are given in the Vega system and the $r$-band magnitudes -- in the AB system.}
\label{fig:opt-lc-BRrG}
\end{figure}

We obtained seventeen non-interrupted light curves  of the \fgl\ counterpart candidate that cover time spans 
of up to 4 hours per night with the time resolution $\sim$200--300 s 
 in different bands between 2022 and 2024. 
The curves obtained during a specific night generally demonstrate irregular brightness variations. 
Examples of the $V$-band curves obtained in 2023 are presented in Fig.~\ref{fig:opt-lc-V}.
As seen, on some nights (e.g., May 31,  September 14 and  16), the source was observed in a relatively stable state   
with $V\approx 20$ mag accompanied by 10--20 min flickering 
with an amplitude  of $\Delta V \approx 0.3$--0.4 mag. 
However, on some other nights (June 1 and September 15) it demonstrated relatively strong flares, with  $\Delta V \approx 1$ mag and    
$\approx$1 h duration.  
Similar behaviour was observed at other epochs 
in different bands (Fig.~\ref{fig:opt-lc-BRrG}).  
For instance, on May 23, 2023 the source showed an  
$\approx$0.5 h long flare in the $B$ and $R$ bands, and a 
flickering before and after it, while on July 26, 2024 the source was relatively stable with a flickering of $\lesssim$0.3 mag, 
 as seen from the light curve in the $G$ band. 

The amplitudes of the brightness variations in our data are generally  consistent 
with the range of the variability found in the sparse 
ZTF and \ps\ data (Fig.~\ref{fig:lc-cat}). 
At the same time, using the Lomb-Scargle periodogram method \citep{lomb,scargle} 
we found no significant periodic brightness modulation 
in our less sparse and noisy data within the period search range from 30 min to 24 h.

\subsection{Optical spectra}
\label{subsec:opt-spec}

The processing of the RTT-150 spectral data was performed  using the DECH software \citep{Galazutdinov22}.
To obtain the source flux densities, the spectra of the spectrophotometric standard stars BD17d4708 and BD26d2606 were obtained during the same nights as the target. 

The \gem\ data processing  was performed  using the DRAGONS package \citep{DRAGONS}.
To calibrate the flux, spectra of the spectrophotometric standard  EG131 were obtained during the same night as the target.

\begin{figure*}[h]
\begin{center}
\begin{minipage}{1.0\linewidth}
\centering{\includegraphics[width=\linewidth]{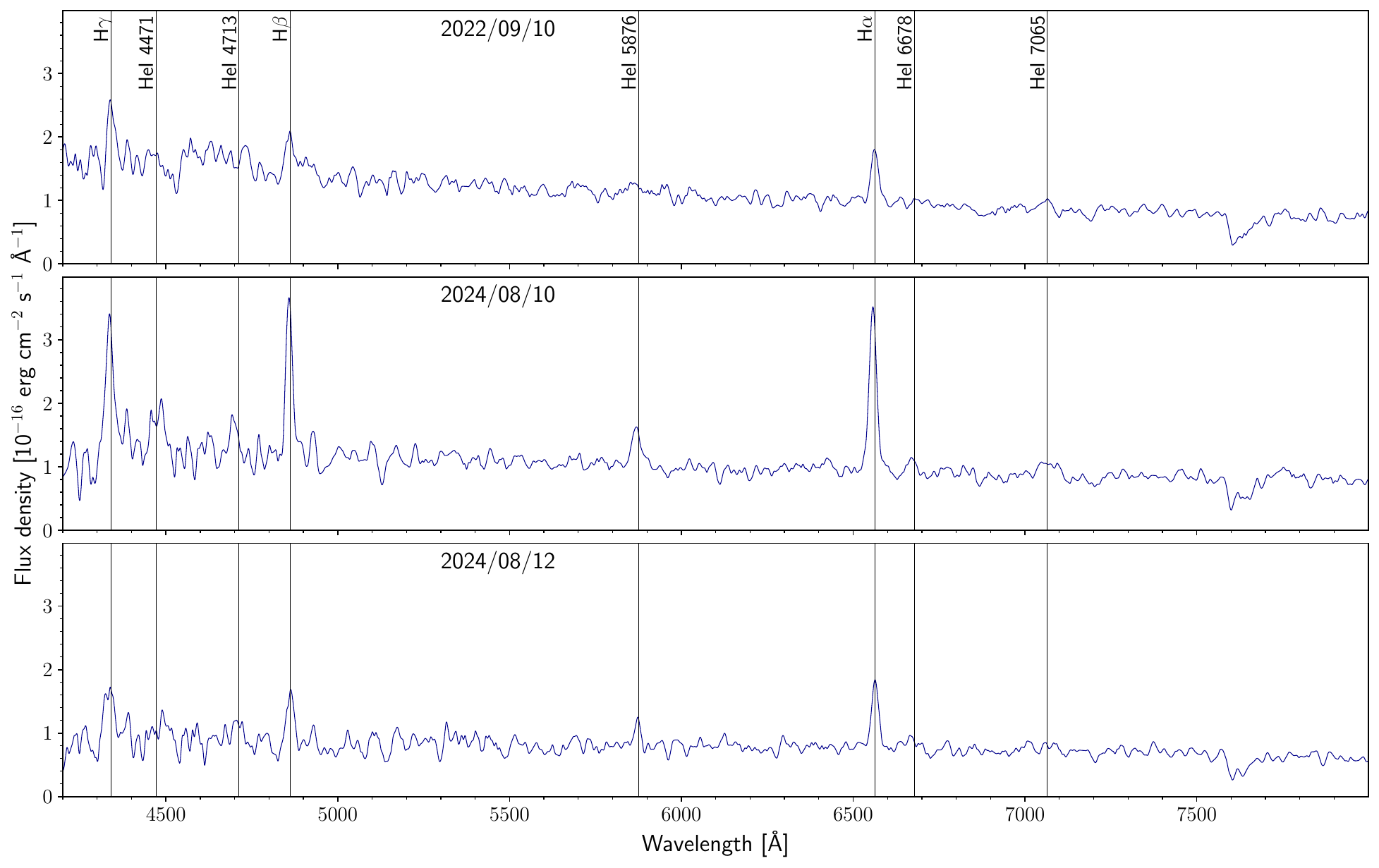}}
\end{minipage}
\end{center}
\caption{Unabsorbed optical spectra of the likely optical counterpart to \fgl\ obtained with RTT-150 on different dates.
Positions of H and He lines are shown. }
\label{fig:opt-spec-RTT}
\end{figure*}

Optical spectra  of \fgl\  demonstrate a flat nonstellar continuum with hydrogen Balmer 
and Paschen  and He~I 
emission lines.  
Typical examples of spectra obtained with  RTT-150 are presented in Fig.~\ref{fig:opt-spec-RTT}.
They are corrected for the interstellar reddening $E(B-V) \approx 0.14$ mag which is the maximal value in the source direction according to the 3D dust map by \citet{dustmap2019}. The reddening quickly increases with the distance and reaches its maximum at about 1.45 kpc, which is close to the lower limit on the distance of the source. 
As seen from these figures, 
within the spectral resolution of RTT-150,  
the H~I and He~I lines are single peaked, 
their intensities vary in time by a factor of two,  
and no significant He~II emission and/or  
any absorption features are resolved. 
The profiles of Balmer lines in the spectrum obtained on August 10, 2024, when they were the brightest, are blue-shifted by about 8~\AA\ with respect to the rest of the spectra.

\begin{figure*}[h]
\begin{center}
\begin{minipage}{1.0\linewidth}
\centering{\includegraphics[width=\linewidth]{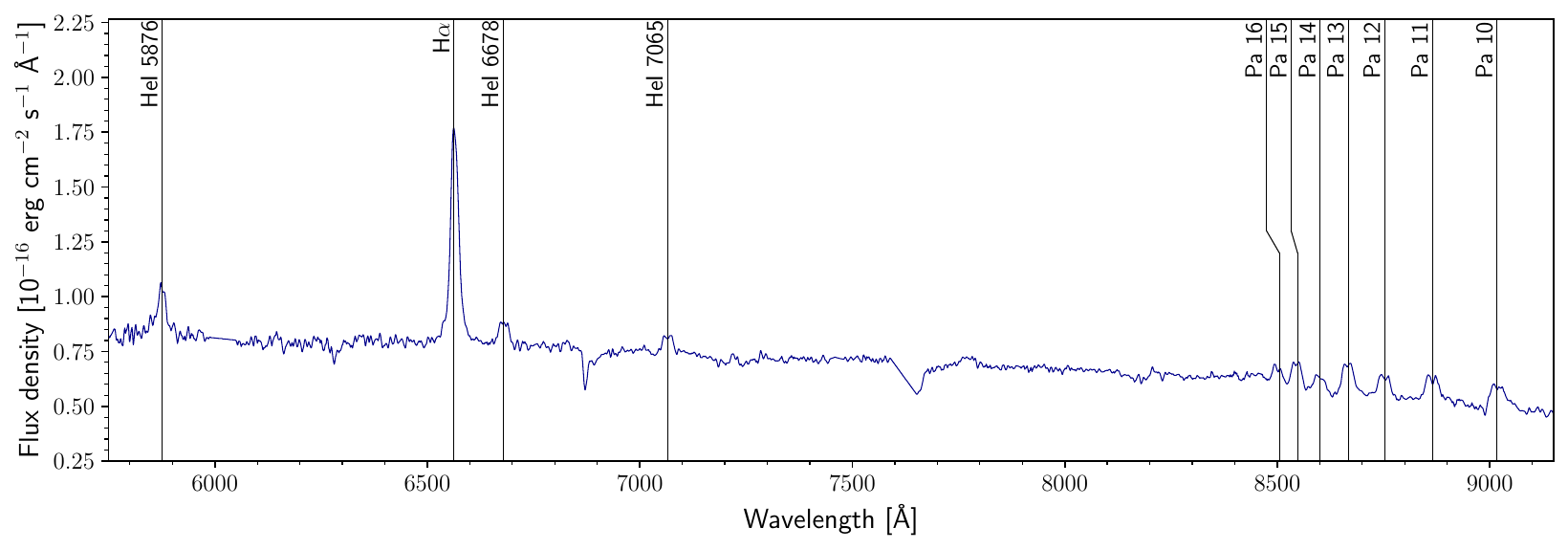}}
\put(-290,85){\includegraphics[width=0.33\linewidth]{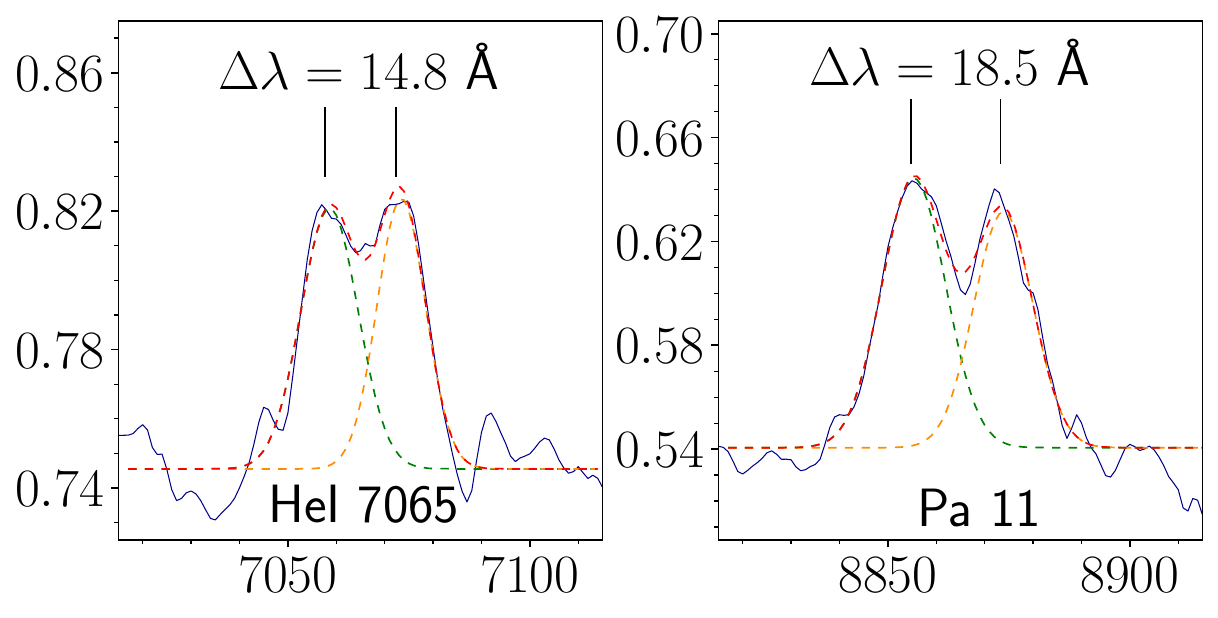}}
\end{minipage}
\end{center}
\caption{Unabsorbed optical spectrum of the likely optical counterpart to \fgl\ obtained with the \gem-North.
Positions of H and He lines are shown.
In the insets, the zoomed-in regions around He~I~7065 and Pa 11 lines are presented.
The separations between the two peaks are marked.
The best-fitting two-component Gaussian models are shown by the dashed red lines, while their components  -- by the green and orange dashed lines.}
\label{fig:opt-spec-Gemini}
\end{figure*}

The combined spectrum obtained with \gem\ is shown in Fig.~\ref{fig:opt-spec-Gemini}. It has a higher quality and spectral resolution, 
as compared to that from the RTT-150, and allows one to see the spectrum of the source at longer wavelengths,          
where the double-peaked structure of 
He~I~6678, 7065 \AA\  and hydrogen Paschen lines is resolved.   
At the same time, H$\alpha$ and 
possibly He~I~5876 emission lines remain single-peaked 
as in the RTT-150 spectra, probably due to low spectral 
resolution in the bluer part of the spectrum.
As an example, the distances between two  peaks  of the He~I~7065 \AA\ and  Pa 11 emission lines are about 14.8~\AA\ and 18.5~\AA, respectively (both correspond to $\approx$630 km~s$^{-1}$). They were calculated from the difference of the centroids of the two-component Gaussian profile used to fit the observed lines.
The full width at half maximum (FWHM) of the H$\alpha$ line is about 960 km~s$^{-1}$.
We note that He~II~4686 line is not detected in the \gem\ spectrum, likely due to the low efficiency below $\approx$4800 \AA\ and/or its weakness.

\subsection{X-ray spectra and variability}
\label{sec:xray-spec}
\begin{figure}[t]
\begin{center}
\begin{minipage}{1.0\linewidth}
\includegraphics[width=1.\linewidth]{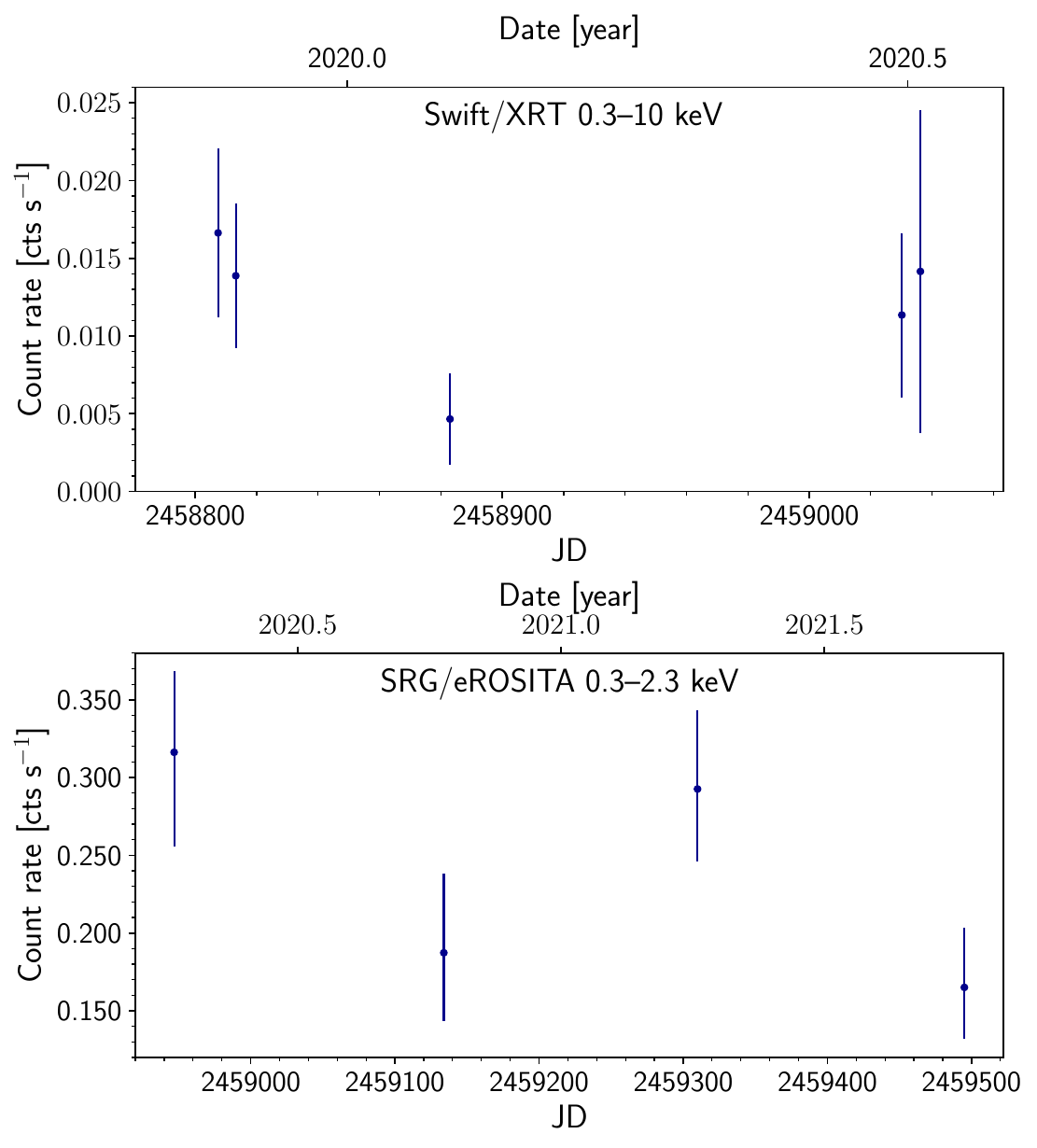}
\end{minipage}
\end{center}
\caption{X-ray light curves of the likely counterpart to \fgl\ obtained with \sw/XRT and SRG/\eros.}
\label{fig:xray-lc}
\end{figure}


\begin{figure}[h]
\begin{center}
\begin{minipage}{1.0\linewidth}
\includegraphics[width=1.0\linewidth]{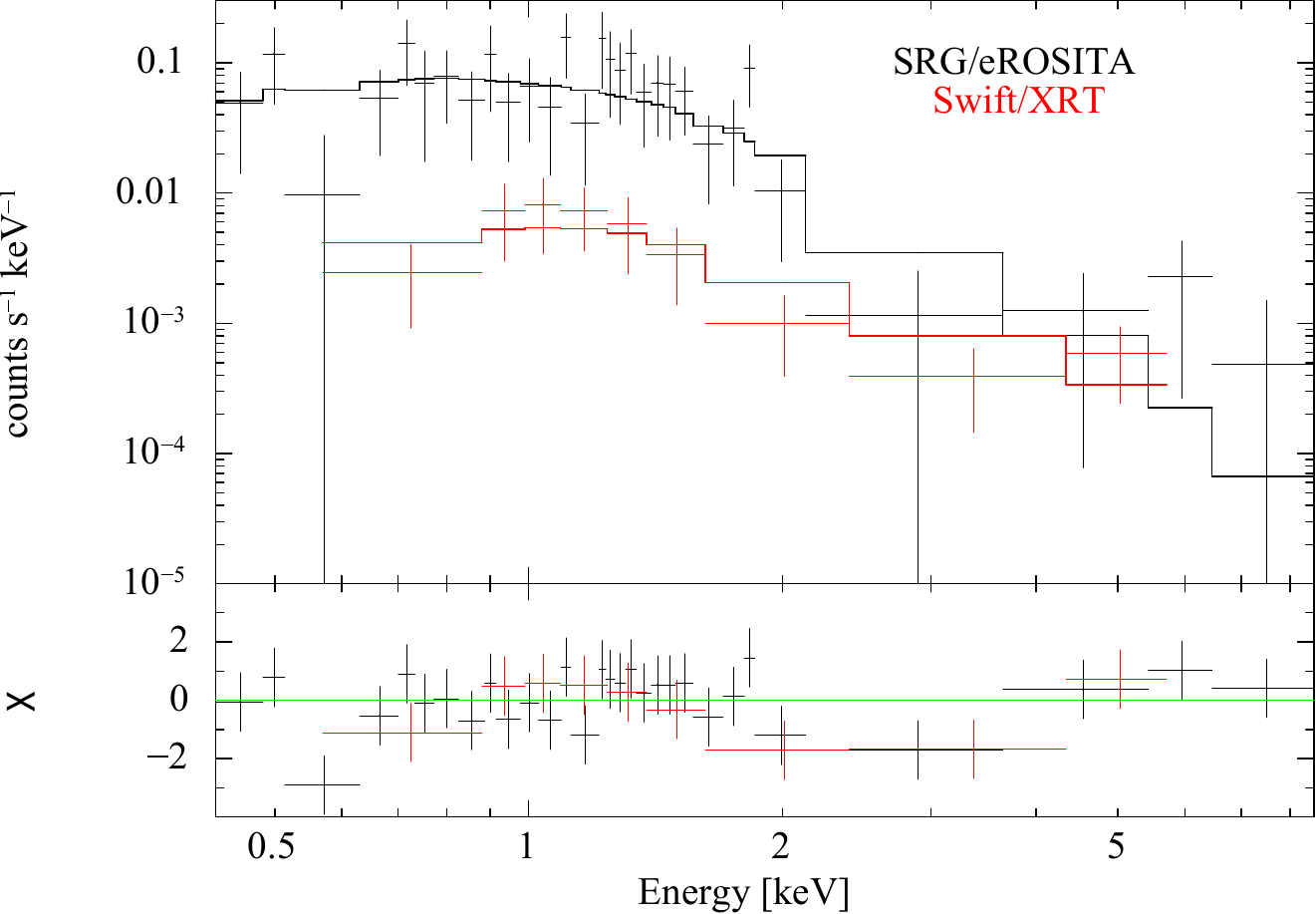}
\end{minipage}
\end{center}
\caption{X-ray spectrum of the \fgl\ likely counterpart obtained with \eros\ and \sw, the best-fitting PL model (top panel) and residuals (bottom panel). For illustrative purposes, the spectra were grouped to ensure at least 3 counts per bin.}
\label{fig:xray-spec}
\end{figure}

We extracted spectra and light curves of the source from the \sw/XRT and SRG/\eros\  data. 
For the \sw\ data, we used the \sw-XRT data products generator\footnote{\url{https://www.swift.ac.uk/user\_objects/}} \citep{evans2009}. 
The \eros\ source counts were 
extracted from a 60\asec-radius circle.  
For the background, we used an annulus region with inner and outer radii of 150\asec\ and 300\asec\ around the source. 

No significant X-ray variability of the \fgl\ counterpart candidate was detected with \sw/XRT and \eros: mean count rates in different observations are compatible with each other at a $\lesssim$2$\sigma$ level\footnote{One \sw\ data set with a very short  exposure time of  $\approx$53 s  provides only an uninformative 3$\sigma$ upper limit on the count rate of 0.2 cts~s$^{-1}$.} (see Fig.~\ref{fig:xray-lc}). 
The low number and short duration of the observations together with the faintness of the source prevented the analysis of the count rate bimodality which is a characteristic feature of tMSPs.

Since the source variability is not significant, we analysed the time integrated spectra. 
As a result, we obtained 28 and 83 net counts in total for the \sw\ (0.3--10 keV band) and \eros\ (0.3--9 keV band) data, respectively.
Both spectra were binned to ensure $\geq 1$ counts per energy bin.
We fitted the spectra with the X-Ray Spectral Fitting Package (XSPEC) v.12.13.1 \citep{xspec1996}.
Due to the low number of counts, we used $C$-statistic \citep{cash}.
To account for interstellar absorption, we applied the \texttt{tbabs} model and \texttt{wilm} abundances \citep{wilms2000} and the maximum reddening value mentioned above. 
Using the empirical relation from \citet{foight2016}, we converted $E(B-V)$ to the absorption column density, obtaining $N_{\rm H} = 1.2 \times 10^{21}$~cm$^{-2}$. 
We fixed this value during the fitting procedure\footnote{In addition, we checked the interstellar absorption using the 3D--\nh\ tool by \citet{Doroshenko2024} and obtained $N_{\rm H} < 4 \times 10^{21}$~cm$^{-2}$. This is compatible with the value used in the fitting procedure.}.

We found that the spectra can be well described by the power-law (PL) model, and their photon indices and model normalisations are in agreement within 1$\sigma$ uncertainties. 
Thus, we fitted the spectra simultaneously with common parameters, 
resulting in the photon index $\Gamma = 1.74\pm0.19$, the unabsorbed flux of $F_X = 5.8^{+1.2}_{-1.0} \times 10^{-13}$~\ergscm\ 
in the 0.5--10 keV band and $C=83$ for 111 degrees of freedom (the uncertainties correspond to 1$\sigma$ confidence intervals).
The spectra with the best-fitting model are shown in Fig.~\ref{fig:xray-spec}.


\section{Discussion and conclusions}
\label{sec:discussion}

We found a likely X-ray/optical counterpart to the unassociated $\gamma$-ray source \fgl.
The counterpart is definitely located in the Galaxy since it has a significant proper motion of 18.5 mas~yr$^{-1}$ and the distance is estimated to be in a range of 1.3--3.8 kpc. 

The source magnitudes in the \gaia\ catalogue are $G = 19.3$, $G_{\rm BR} = 19.97$ and $G_{\rm RP} = 18.54$, which gives the colour $G_{\rm BR}-G_{\rm RP} = 1.43$ mag.
Using its X-ray flux and $G$-band magnitude, we estimated the X-ray-to-optical flux ratio $F_X$/$F_{\rm opt} \sim 3$. 
This value, together with the optical colour, corresponds to classes of accreting compact objects (see fig.~1 in \citealt{Rodriguez2024}) which include cataclysmic variables (CVs), super-soft sources (SSS), LMXBs, high-mass X-ray binaries (HMXBs), and binary MSPs (BWs and RBs, including tMSPs).

The presence of Balmer hydrogen and He emission lines in the optical spectrum of the \fgl\ likely counterpart supports the interpretation of the source as an accreting system. In the case of an accretion disk, lines usually show double-peaked profiles due to the Doppler motions of the matter within the binary.
This is seen in the spectra obtained with \gem.
The blueshift of about 8~\AA\ ($\approx$300 km~s$^{-1}$) seen in one of the RTT-150 spectra can be caused either by the orbital motion or accretion outflow. 
Spectroscopy with high time resolution is necessary to clarify that.

Investigation of the long-term variations of the source (Fig.~\ref{fig:lc-cat}) does not show any presence of prolonged low-emission states that are usually observed in polars and intermediate polars \citep[e.g.][]{mason&santana2015, covington2022,Kolbin2024}.
We did not find periodic variations in the optical data. 
The source short-term behaviour is rather complex: the relatively stable states with $V\approx20$ mag are accompanied by flickering and flares of various amplitudes ($\sim$0.4--1 mag) and durations ($\sim$10 min--2 h). 
Such rapid optical variability is different from that of most CVs, but is typical for tMSP \citep{papitto&demartino2022}.

\begin{figure}[h]
\begin{center}
\begin{minipage}{1.0\linewidth}
\includegraphics[width=1.0\linewidth]{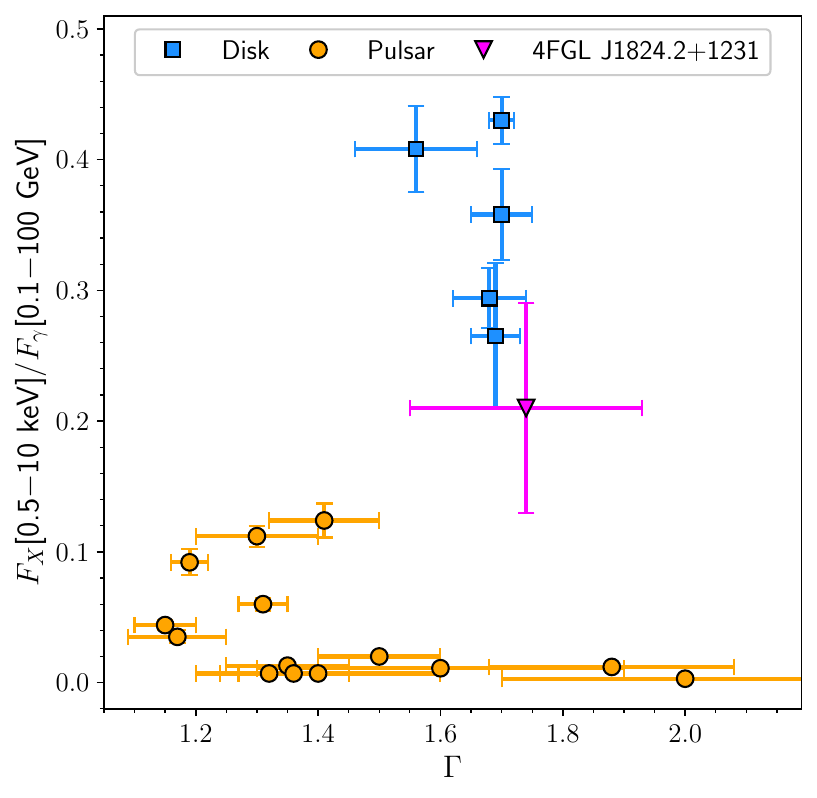}
\end{minipage}
\end{center}
\caption{Ratio of X-ray to $\gamma$-ray flux vs X-ray photon index for tMSPs or their candidates in the subluminous disk state (blue squares) and RBs or tMSPs in the pulsar regime (orange circles; adopted from \citealt{miller2020}). 
The position of \fgl\ is shown by the magenta triangle.}
\label{fig:gamma-flux-ratio}
\end{figure}

If we assume that the object is a real counterpart of the \fermi\ source \fgl, then its X-ray (0.5--10 keV) and $\gamma$-ray (0.1--100 GeV) luminosities 
are $L_X = 6.9^{+1.5}_{-1.2}\times 10^{31}\ D^2_{\rm kpc}$~\ergs\ and $L_\gamma = (3.2 \pm 1.1)\times 10^{32}\ D^2_{\rm kpc}$~\ergs, where $D_{\rm kpc}$ is the distance in units of kpc.
For the distance range 1.3--3.8 kpc mentioned above, this yields $L_X \approx 10^{32}$--$10^{33}$ \ergs\ and $L_\gamma \approx (0.36-6.2)\times10^{33}$~\ergs.
The X-ray-to-$\gamma$-ray flux ratio $F_X/F_\gamma=0.21\pm0.08$
and the derived photon index of the X-ray spectrum $\Gamma\approx1.7$ are typical for tMSPs in the sub-luminous disk state (see Fig.~\ref{fig:gamma-flux-ratio} and \citealt{papitto&demartino2022}). 
The X-ray luminosity range is also compatible with this class, which shows $\sim$10$^{33}$--10$^{34}$ \ergs. 
This supports the tMSP nature of \fgl.

We can exclude the LMXB nature of the source since its X-ray spectrum with the photon index $\Gamma \approx 1.7$ is harder than those of LMXBs in the luminosity state with $L_X \sim 10^{32}$--10$^{33}$ \ergs\ \citep{tanaka1997}. It is also harder than SSS spectra \citep{sss}.
The source is too faint to be an HMXB with large and bright OB-type donors with absolute magnitudes of $M_V<0$ \citep[e.g.][]{wegner2000}. 

On the other hand, based on the optical and X-ray properties, we cannot completely exclude the possibility that the source belongs to the CV family. 
If so, its association with the $\gamma$-ray source \fgl\ is unlikely since only few CVs were detected in $\gamma$-rays. 
These are novae eruptions \citep[e.g.][]{sokolovsky2022,sokolovsky2023} and possibly two rapidly rotating magnetic CVs, AE Aqr and AR Sco \citep{Meintjes2023}.
We also note that the two confirmed tMSPs, PSR J1023+0038 and PSR J1227$-$4853, were initially classified as CVs \citep{bond2002,Masetti2006} and only further observations revealed their real nature. 

To conclude, \fgl\ is a promising tMSP candidate.
To firmly establish its nature, new observations are needed.
Observing the transition to the pulsar state would definitely classify
\fgl\ as a tMSP. 
Optical monitoring is the most useful tool for achieving this goal.
Detection of sharp transitions between the high and low brightness modes in X-rays or in the optical/UV would be also valuable as a unique marker of tMSPs.

When this work was in the final stage of preparation, a manuscript by \citet{kyer2025}
containing an independent multiwavelength study  
of this source was accepted for publication. 
Using their optical and X-ray data, the authors obtained results compatible with ours and concluded that the properties of \fgl\ are consistent with the 
observed properties of the subluminous disk state of tMSPs. Together, the results of both independent studies reinforce this conclusion.


\begin{acknowledgements}
We thank the anonymous referee for useful comments.
The work is based upon observations carried out at the Observatorio Astron\'omico Nacional on the Sierra San Pedro M\'artir (OAN-SPM), Baja California, M\'exico. We thank the daytime and night support staff at the OAN-SPM for facilitating and helping obtain our observations. 
The authors are grateful to TUBITAK, the Space Research
Institute, the Kazan Federal University, and the
Academy of Sciences of Tatarstan for their partial
support in using RTT-150 (the Russian–Turkish
1.5-m telescope in Antalya).
This work has made use of data from the European Space Agency (ESA) mission {\it Gaia} (\url{https://www.cosmos.esa.int/gaia}), processed by the {\it Gaia} Data Processing and Analysis Consortium (DPAC, \url{https://www.cosmos.esa.int/web/gaia/dpac/consortium}). Funding for the DPAC has been provided by national institutions, in particular the institutions participating in the {\it Gaia} Multilateral Agreement.
The Pan-STARRS1 Surveys (PS1) and the PS1 public science archive have been made possible through contributions by the Institute for Astronomy, the University of Hawaii, the Pan-STARRS Project Office, the Max-Planck Society and its participating institutes, the Max Planck Institute for Astronomy, Heidelberg and the Max Planck Institute for Extraterrestrial Physics, Garching, The Johns Hopkins University, Durham University, the University of Edinburgh, the Queen's University Belfast, the Harvard-Smithsonian Center for Astrophysics, the Las Cumbres Observatory Global Telescope Network Incorporated, the National Central University of Taiwan, the Space Telescope Science Institute, the National Aeronautics and Space Administration under Grant No. NNX08AR22G issued through the Planetary Science Division of the NASA Science Mission Directorate, the National Science Foundation Grant No. AST-1238877, the University of Maryland, Eotvos Lorand University (ELTE), the Los Alamos National Laboratory, and the Gordon and Betty Moore Foundation.
Based on observations obtained with the Samuel Oschin Telescope 48-inch and the 60-inch Telescope at the Palomar Observatory as part of the Zwicky Transient Facility project. ZTF is supported by the National Science Foundation under Grants No. AST-1440341 and AST-2034437 and a collaboration including current partners Caltech, IPAC, the Oskar Klein Center at Stockholm University, the University of Maryland, University of California, Berkeley , the University of Wisconsin at Milwaukee, University of Warwick, Ruhr University, Cornell University, Northwestern University and Drexel University. Operations are conducted by COO, IPAC, and UW.
This work used data obtained with eROSITA telescope onboard SRG observatory. The SRG observatory was built by Roskosmos in the interests of the Russian Academy of Sciences represented by its Space Research Institute (IKI) in the framework of the Russian Federal Space Program, with the participation of the Deutsches Zentrum für Luft- und Raumfahrt (DLR). The SRG/eROSITA X-ray telescope was built by a consortium of German Institutes led by MPE, and supported by DLR.  The SRG spacecraft was designed, built, launched and is operated by the Lavochkin Association and its subcontractors. The science data are downlinked via the Deep Space Network Antennae in Bear Lakes, Ussurijsk, and Baykonur, funded by Roskosmos. The eROSITA data used in this work were processed using the eSASS software system developed by the German eROSITA consortium and proprietary data reduction and analysis software developed by the Russian eROSITA Consortium.
DAZ thanks Pirinem School of Theoretical Physics for hospitality. 
The work of DAZ, AVK and YAS (analysis of archival and OAN-SPM optical data and analysis of X-ray data) was supported by the baseline project FFUG-2024-0002 of the Ioffe Institute. AYK acknowledges the DGAPA-PAPIIT grant IA105024. 
The work of IFB, ENI, MAG, MVS was supported through subsidy FZSM-2023-0015 of the Ministry of Education and Science of the Russian Federation allocated to the Kazan Federal University for
the State assignment in the sphere of scientific activities.
\end{acknowledgements}

%
\bibliographystyle{aa} 
\bibliography{arxiv-j1824} 
%

\end{document}